\documentclass[runningheads]{cl2emult}

\usepackage{makeidx}  
\usepackage{graphicx} 
\usepackage{subeqnar} 
\usepackage{multicol} 
\usepackage{cropmark} 
\usepackage{physbb}   
\makeindex            

\begin{document}
\title*{General Dynamic Wormholes and Violation of the Null Energy
Condition}
\titlerunning{Dynamic Wormholes}
%
\author{David Hochberg\inst{1}
\and Matt Visser\inst{2}}
\authorrunning{David Hochberg and Matt Visser}
%
%
\institute{Laboratorio de Astrof\'\i sica Espacial y F\'\i sica Fundamental\\
Apartado 50727, Madrid, Spain\\
\and Physics Department, Washington University\\
     St. Louis, Missouri 63130-4899, USA}

\maketitle              

\begin{abstract}
\index{abstract} Although wormholes can be treated as topological 
objects in spacetime and from a global point-of-view, a
precise definition of what a wormhole throat is and where 
it is located can be 
developed and treated entirely in terms of local geometry. 
This has the advantage
of being free from unnecesary technical assumptions about 
asymptotic flatness,
and other global properties of the spacetime containing 
the wormhole. 
We discuss our recent work proving that the violation 
of the null energy
condition (NEC) is a {\em generic} feature of all wormholes, 
whether they be time-dependent
or static,
and demonstrate 
that time-dependent
wormholes have {\em two} throats, one for each direction through
the wormhole, which coalesce only in the static limit. 
\end{abstract}

\section{Introduction}
The fact that traversable wormholes are
accompanied by unavoidable violations of the null energy condition
(NEC) is perhaps one of the most important aspects of Lorentzian wormhole
physics~\cite{Morris-Thorne,MTY,Visser-Book}. The original proof of
the necessity for NEC violations at or near the throat of a
traversable wormhole was limited to the static spherically symmetric
Morris-Thorne wormhole~\cite{Morris-Thorne}, though it was soon after
realized that NEC violations typically occurred in at least some
explicit examples of static non-symmetric~\cite{Visser-Examples} and
spherically-symmetric time-dependent~\cite{Visser-Surgery}
wormholes. A considerably more general proof of the necessity of NEC
violations was provided by the {\em topological censorship theorem} of
Friedman, Schleich, and Witt~\cite{FSW} though this theorem requires
many technical assumptions concerning asymptotic flatness and
causality conditions that limit its applicability.

We have recently adopted a different strategy by developing
new general theorems concerning energy condition violations at and
near the throat of traversable wormholes \cite{Hochberg-Visser,Visser-Hochberg},
by focusing attention only on
the local behavior of the geometry at and near the throat, and
dispensing with all assumptions about symmetry, asymptotic behaviour, and
causal properties. This strategy was inspired by the fact that there
are many classes of spacetime configurations 
that we would meaningfully wish to call
wormholes that possess either 
trivial topology~\cite{Visser-Book,Hochberg-Kephart} 
or do
not necessarily possess asymptotically flat
regions~\cite{HPS}. The strategy 
that we have developed views a wormhole throat as a marginally anti-trapped
surface \cite{Hochberg-Visser2,Hochberg-Visser3} 
that is, a closed two-dimensional spatial hypersurface
such that one of the two future-directed null geodesic 
congruences orthogonal to it is just beginning to diverge. 
Physically, this definition relies only on the properties of bundles
of light rays passing through the region surrounding the throat, and not
on how that throat may or may not embed in some fictitious 
Euclidean embedding
space.

\section{Definition of generic wormhole throats}

For the generic but {\em static} case, the throat was defined
as a two-dimensional hypersurface of minimal
area~\cite{Hochberg-Visser,Visser-Hochberg}.  The time independence
allows one to locate that minimal hypersurface entirely within one
of the constant-time three-dimensional spatial slices, and the
conditions of extremality and minimality can be applied and enforced
within that single time-slice. For a static throat, variational
principles involve performing arbitrary time-independent surface
deformations of the hypersurface in the remaining spatial direction
orthogonal to the hypersurface, which can always be taken to be
locally Gaussian. By contrast, in the time-dependent case, it may
not be possible to define the throat by working within one time slice:
the dynamic throat is an extended object in spacetime, and the
variational principle must be carried out employing surface
deformations in the two independent {\em null} directions orthogonal
to the hypersurface: say, $\delta u_+$ and $\delta u_{-}$.  This,
by the way, demonstrates why it is that {\em in general} the 
embedding of the spatial
part of a wormhole spacetime in an Euclidean ${\bf R}^n$ is no
longer a reliable operational technique for defining ``flare-out''
in the time-dependent case. We will come back to this point below.  
Of course, in the static limit these
two variations will no longer be independent and arbitrary deformations
in the two null directions reduce to a single variation in the
constant-time spatial direction as demonstrated below.  

\subsection{Geometric Preliminaries}

We now set up and define the properties of throats
in terms of the null congruences. Bear in mind that a throat will
be characterized in terms of the behavior of a single set of null
geodesics orthogonal to it.  We define a wormhole throat $\Sigma_{u
+}$ (there is also one for the other null congruence) to be a closed
2-dimensional hypersurface of minimal area taken in one of the
constant-$u_{+}$ slices, where $u_{+}$ is an affine parameter
suitable for parameterizing the future-directed null geodesics
$l_+$ orthogonal to $\Sigma_{u +}$.   All this means is that we
imagine ``starting'' off a collection of light pulses along the
hypersurface and we can always arrange the affine parameterizations
of each pulse to be equal to some constant on the hypersurface; we
take this constant to be zero.  We wish to emphasize that there is
a corresponding definition for the other throat $\Sigma_{u-}$. In
the following, we define and develop the conditions that both
hypersurfaces must satisfy individually to be considered as throats,
and shall do so in a unified way by treating them together by
employing the $\pm$-label.  Our next task is to compute the
hypersurface areas and impose the conditions of extremality and
minimality directly and to express these constraints in terms of
the expansion $\theta$ of the null geodesics. The area of 
the two-dimensional spatial hypersurface $\Sigma_{u\pm}$
is given by
\begin{equation}
\label{E:area-v}
A(\Sigma_{u\pm}) = \int_{\Sigma_{u\pm}} \sqrt{\gamma} \, d^2x.
\end{equation}

An arbitrary variation of the surface with respect to deformations
in the null direction parameterized by $u_{\pm}$ is 
\begin{eqnarray}
\label{E:firstvary}
\delta A (\Sigma_{u\pm}) & = & \int_{\Sigma_{u\pm}}
 \frac{d \sqrt{\gamma}}{d u_{\pm}}
\,\delta u_{\pm}(x) 
\, d^2x.\nonumber \\
&=& \int_{\Sigma_{u\pm}} \sqrt{\gamma} \, \frac{1}{2}\gamma^{ab}\,
\frac{d \gamma_{ab}}{d u_{\pm}} \, \delta u_{\pm}(x) 
\, d^2x.
\end{eqnarray}
If this is to vanish for arbitrary variations $\delta u_{\pm}(x)$, then
we must have that 
\begin{equation}\label{extremal}
\frac{1}{2}\gamma^{ab}
\frac{d \gamma_{ab}}{d u_{\pm}} = 0,
\end{equation}
which expresses the fact that the hypersurface 
$\Sigma_{u \pm}$ is extremal.

This condition of hypersurface extremality can also be phrased
equivalently and directly in terms of the expansion of the null
congruences.  The simplest way to do so is to consider the Lie
derivative ${\cal L}^{\pm}_l$ acting on the full spacetime metric:
\begin{eqnarray}\label{Lie}
{\cal L}^{\pm}_l g_{ab} 
&=& l^c_{\pm}\nabla_c g_{ab} + g_{cb}\nabla_al^c_{\pm} 
+ g_{ac}\nabla_b l^c_{\pm}\nonumber \\
&=& \nabla_a l_{\pm b} + \nabla_b l_{\pm a}\nonumber \\
&=& {B^{\pm}}_{ba} + {B^{\pm}}_{ab} = 2{B^{\pm}}_{(ab)},
\end{eqnarray}
with the second equality holds due to the covariant constancy
of the metric. The third line defines the tensor field $B_{ab}$ 
as the covariant derivative of
the future-directed null vectors (there is one such tensor field for each
null congruence):
\begin{equation}
\label{E:B}
B^{\pm}_{ab} \equiv \nabla_b l_{\pm a},
\end{equation}

We now use the decomposition (\ref{E:decomposition}) of the spacetime
metric 
following the description of Carter~\cite{Carter}. The 
future-directed ``outgoing'' null vector $l_{+}^a$ and future-directed
``ingoing'' null vector $l_{-}^a$ introduced above 
together with a spatial orthogonal projection
tensor $\gamma^{ab}$ can be chosen satisfying the following relations:
\begin{eqnarray}
\label{E:properties}
l^a_{+}l_{+a} &=& l^a_{-}l_{-a} = 0 ,\qquad
l^a_{+}l_{-a} = l^a_{-}l_{+a} = -1 \nonumber \\
l^a_{\pm}\gamma_{ab} &=&  0 , \qquad
\gamma^a_c\gamma^{cd} = \gamma^{ad}.
\end{eqnarray}
In terms of these null vectors and projector, we can decompose the
full spacetime metric (indeed, any tensor) uniquely:
\begin{equation}
\label{E:decomposition}
g_{ab}= \gamma_{ab} - l_{-a}l_{+b} - l_{+a}l_{-b}.
\end{equation}
Physically, this decomposition leads to a parameterization of
spacetime points in terms of two spatial coordinates (typically
denoted $x$) plus two null coordinates [$u_\pm$, or sometimes
$(u,v)$]. (We do not want to prejudice matters by taking the
words ``outgoing'' and ``ingoing'' too literally, since outside and
inside do not necessarily make much sense in situations of nontrivial
topology. The critical issue is that the spacelike hypersurface
must have two sides and $+$ and $-$ are just two convenient labels
for the two null directions.)

Using (\ref{Lie}) we can now work out the Lie derivative 
using the Leibnitz rule:
\begin{eqnarray}
{B^{\pm}}_{(ab)} 
&=& \frac{1}{2}{\cal L}^{\pm}_l g_{ab} \nonumber \\
&=& \frac{1}{2}{\cal L}^{\pm}_l 
(\gamma_{ab} -l_{-a}l_{+ b} -l_{+ a} l_{- b}),\nonumber \\
&=& \frac{1}{2}{\cal L}^{\pm}_l \gamma_{ab} 
-\frac{1}{2}(l_{-a} {\cal L}^{\pm}_ll_{+b}
 + l_{+b} {\cal L}^{\pm}_l l_{-a} + (a \leftrightarrow b) ),
\end{eqnarray}
from which, and using the properties in (\ref{E:properties}), implies
\begin{eqnarray}
\theta_{\pm} &\equiv&  g^{ab}{B^{\pm}}_{(ab)} =
\gamma^{ab}{B^{\pm}}_{(ab)} \nonumber \\
       &=& \frac{1}{2}\gamma^{ab}{\cal L}^{\pm}_l \gamma_{ab}
       = \frac{1}{2}\gamma^{ab}\frac{d \gamma_{ab}}{d  u_{\pm}}.
\end{eqnarray}
Note the trace of the symmetrized tensor $B_{(ab)}$ defines
the divergence
$\theta$.
So the condition that the area of the hypersurface be 
extremal (\ref{extremal}) is
simply that the expansion of the null geodesics vanish at the
surface: $\theta_{\pm} = 0$.  

\subsection{Flare-out}

To ensure that the area be {\em minimal},
we need to impose an additional constraint and shall require that
$\delta^2 A (\Sigma_{u\pm}) \geq 0$. By explicit computation,
\begin{eqnarray}
\label{E:minimality}
\delta^2 A(\Sigma_{u\pm}) &=& \int _{\Sigma_{u\pm}} \sqrt{\gamma} \left(
{\theta_{\pm}}^2 + \frac{d \theta_{\pm}}{d u_{\pm}} \right)
\delta u_{\pm}(x)\, \delta u_{\pm}(x) d^2x \nonumber \\
&=& \int_{\Sigma_{u\pm}} \sqrt{\gamma}\, \frac{d \theta_{\pm}}{d u_{\pm}}
\, \delta u_{\pm}(x)\, \delta u_{\pm}(x)\, d^2x \geq 0,
\end{eqnarray}
where we have used the extremality condition ($\theta_{\pm} = 0$) in
arriving at this last inequality.  For this to hold at the throat
for arbitrary variations $\delta u_{\pm}(x)$, it follows from 
$(\delta u_{\pm}(x))^2
\geq 0$, that we must have
\begin{equation}\label{simpleflare}
\frac{d \theta_{\pm}}{d u_{\pm}} \geq 0,
\end{equation}
in other words, the expansion of the cross-sectional area of the
future-directed null geodesics must be locally increasing at the
throat.  This is the precise generalization of the Morris-Thorne
``flare-out'' condition to arbitrary wormhole throats. This makes
eminent good sense since the expansion is the measure of the
cross-sectional area of bundles of null geodesics, and a positive
derivative indicates that this area is locally increasing or
``flaring-out'' as one moves along the null direction. Note that
this definition is free from notions of embedding 
and ``shape''-functions as well as global features of the spacetime.
So in general, we have to deal with two throats: $\Sigma_{u+}$ such
that $\theta_+ = 0$ and $d \theta_+/d u_+ \geq 0$ and $\Sigma_{u-}$
such that $\theta_{-} = 0$ and $d \theta_{-}/d u_{-} \geq 0$.  We
shall soon see that for static wormholes the two throats coalesce
and this definition automatically reduces to the static case
considered in~\cite{Hochberg-Visser,Visser-Hochberg}. The logical
development reviewed here closely parallels that of the
static case though there are important differences.

The conditions that a wormhole throat be both extremal and minimal
are the simplest requirements that one would want a putative throat
to satisfy and which may be summarized in the following definition
(in the following, the hypersurfaces are understood to be closed
and spatial). Since these definitions hold of course for both
throats, we momentarily drop the distinction and suppress the $\pm$
label.

\subsubsection{Definition: Simple flare-out condition}
%
{\em A two-surface satisfies the ``simple flare-out'' condition if
and only if it is extremal, $\theta=0$, and also satisfies ${d
\theta/d u} \geq 0$.}
The characterization of a generic wormhole throat in terms of the
expansion of the null geodesics shows that any two-surface satisfying
the simple flare-out condition is a {\em marginally anti-trapped
surface}, where the notion of trapped surfaces is a familiar concept
that arises primarily in the context of singularity theorems,
gravitational collapse and black hole physics~\cite{Wald,Hawking-Ellis}.

Generically, we would expect the inequality 
$\delta^2 A(\Sigma_{u})> 0$ to be strict, so that 
the surface is truly a minimal (not
just extremal) surface.  This will pertain provided the inequality
${d \theta/d u} > 0$ is a strict one for at least {\em some} points
on the throat.  This suggests the following definition.

\subsubsection{Definition: Strong flare-out condition}
%
{\em A two-surface satisfies the ``strong flare-out'' condition at
the point $x$ if and only if it is extremal, $\theta=0$,  satisfies
$\frac{d \theta}{d u} \geq 0$ everywhere on the surface and if at
the point $x$, the inequality is strict:}
\begin{equation}\label{strongflare}
\frac{d \theta}{d u} > 0.
\end{equation}
If the latter strict inequality holds for all $x \in \Sigma_u$ in
the surface, then the wormhole throat is seen to correspond to a
{\em strongly anti-trapped surface}. 
It is sometimes sufficient and convenient to work with a weaker,
integrated forms of the flare-out condition. These are described in
detail in \cite{Hochberg-Visser3}.

\subsection{Static limit}

In a static spacetime, a wormhole throat is a closed two-dimensional
spatial hypersurface of minimal area that, without loss of generality,
can be located entirely within a single constant-time spatial
slice~\cite{Hochberg-Visser,Visser-Hochberg}.  Now, for any static
spacetime, one can always decompose the spacetime metric in a
block-diagonal form as
\begin{equation}
g_{ab} = -V_a V_b + {}^{(3)}g_{ab},
\end{equation}
where $V^a = \exp[\phi](\frac{\partial}{\partial t})^a$ is a timelike
vector field orthogonal to the constant-time spatial slices and
$\phi = \phi(x)$ is some function of the spatial coordinates only. In the
vicinity of the throat we can always set up a system of Gaussian
coordinates $n$ so that
\begin{equation}
{}^{(3)}g_{ab} = n_a n_b + \gamma_{ab},
\end{equation}
where $n^a = (\frac{\partial }{\partial n})^a$, $n^an_a = +1$, and
$\gamma_{ab}$ is the two-metric of the hypersurface.  Putting these
facts together implies that in the vicinity of any static throat
we may write the spacetime metric as
\begin{equation}
g_{ab} = -V_a V_b + n_an_b + \gamma_{ab}.
\end{equation}
But (\ref{E:decomposition}) holds in general, so comparing both metric
representations yields the identity
\begin{equation}
-l^a_{-} l^b_{+} - l^a_{+} l^b_{-} = V^a V^b + n^a n^b,
\end{equation}
and the following (linear) transformation relates the two metric
decompositions and preserves the inner-product relations in
(\ref{E:properties}):
\begin{equation}
l^a_{-} = \frac{1}{2}(V^a + n^a), \,\,\,\,
l^a_{+} = \frac{1}{2}(V^a - n^a).
\end{equation}
Since the throat is static, $\gamma_{ab}$ is time-independent,
hence when we come to vary the area (\ref{E:area-v}) with respect to
arbitrary perturbations in the two independent null directions we find that
\begin{eqnarray}
\frac{\partial \gamma_{ab}}{\partial u_{+}}\,\delta u_{+} &=& \frac{1}{2} 
\left(\exp[\phi]\frac{\partial \gamma_{ab}}{\partial t}\delta t
+ \frac{\partial \gamma_{ab}}{\partial n}\delta n\right) = \frac{1}{2} 
\frac{\partial \gamma_{ab}}{\partial n}\delta n,\nonumber \\
\frac{\partial \gamma_{ab}}{\partial u_{-}}\, \delta u_{-} &=& \frac{1}{2} 
\left(\exp[\phi]\frac{\partial \gamma_{ab}}{\partial t}\delta t
- \frac{\partial \gamma_{ab}}{\partial n}\delta n\right) = -\frac{1}{2} 
\frac{\partial \gamma_{ab}}{\partial n}\delta n.
\end{eqnarray}
Thus the two variations 
in the null directions are no longer independent, and reduce to
taking a single surface variation in the spatial Gaussian direction.
So, $\theta_+ = 0 \Longleftrightarrow \theta_{-} = 0$ at the same
hypersurface, proving that $\Sigma_{u_+} = \Sigma_{u_-}$ in the
static limit, and so static wormholes have only one throat.
A thorough analysis of the geometric structure of the generic
static traversable wormhole can be found
in~\cite{Hochberg-Visser,Visser-Hochberg}.

\section{Constraints on the stress-energy}

With the definition of wormhole throat made precise we now turn to
derive constraints that the stress energy tensor must obey on (or
near) any wormhole throat. The constraints follow from combining
the Raychaudhuri equation \cite{Carter,Wald,Hawking-Ellis} governing the
rate-of-change of the divergence along the null direction
(there is one for the (+)-congruence and one for the (-)-congruence) 
\begin{eqnarray}
\label{E:Raychaudhuri}
\frac{d \theta_{\pm}}{d u_{\pm}}= -\frac{1}{2}{\theta_{\pm}}^2 &-& 
{\sigma^{\pm}}^{ab}\sigma_{\pm ab}
+ {\omega^{\pm}}^{ab}\omega_{\pm ab} 
- R_c^d l^c_{\pm} l_{\pm d}, 
\end{eqnarray}
with the flare-out
conditions (\ref{simpleflare}) or (\ref{strongflare}). 
We can use the Einstein equation 
($R_{ab}-\frac{1}{2}g_{ab}R = 8\pi T_{ab}$) to cast this
into an equation involving the stress-energy.
Here, $\sigma_{ab}$ and $\omega_{ab}$ denote the symmetric
shear and antisymmetric twist of the null congruence. They are
purely spatial tensors. 
It is clear that these constraints apply with equal validity at
both the $+$ and $-$ throats, and in the following we cover both
classes simultaneously and without risk of confusion by dropping
the $\pm$-labels.

Since all throats are extremal hypersurfaces $(\theta = 0)$ 
the Raychaudhuri equation (\ref{E:Raychaudhuri}) evaluated at the throat
reduces to
\begin{equation}
\label{E:vthroat}
\frac{d\theta}{d u} + \sigma^{ab}\sigma_{ab} = -8\pi T_{ab} \; l^al^b,
\end{equation}
where we have used the Einstein equation
and the fact that the 
null geodesic congruences are hypersurface orthogonal, so that the
twist $\omega_{ab} = 0$ vanishes identically on the throat.  We
make no claim regarding the shear, nor do we need to, 
except to point out that since
$\sigma_{ab}$ is purely spatial, its square $\sigma^{ab}\sigma_{ab}
\geq 0$ is positive semi-definite everywhere (not just on the throat).
Consider a marginally anti-trapped surface, {\em i.e.}, a throat
satisfying the simple flare-out condition. Then the stress energy
tensor on the throat must satisfy
\begin{equation}
T_{ab} \; l^al^b \leq 0.
\end{equation}
The NEC is therefore either violated, or on the verge of being
violated ($T_{ab} \; l^al^b \equiv 0$), on the throat.  Of course,
whichever one of the two null geodesic congruences ($l_{+}$ or
$l_{-}$) you are using to define the wormhole throat (anti-trapped
surface), you must use the {\em same} null geodesic congruence for
deducing null energy condition violations.

For throats satisfying the strong flare-out condition, we have
instead the stronger statement that for all points on the
throat,
\begin{equation}
T_{ab}l^al^b \leq 0, \,\,{\rm and}\,\,
\exists x \in \Sigma_u \,\,\,{\rm such}\,{\rm that}\,\,\, 
T_{ab} \; l^a l^b < 0,
\end{equation}
so that the NEC is indeed violated for at least {\em some} points
lying on the throat.  By continuity, if $T_{ab} \; l^al^b < 0$ at $x$,
then it is strictly negative within a finite open neighborhood of
$x$: $B_{\epsilon}(x)$.  
Finally, for throats that are strongly anti-trapped
surfaces, we derive the most stringent constraint stating that
\begin{equation}
T_{ab}\; l^a l^b < 0 \,\,\, \forall x \in \Sigma_u,
\end{equation}
so that the NEC is violated {\em everywhere} on the throat.

What can we say about the energy conditions in the region surrounding
the throat? This requires knowledge of the expansion, shear and
twist in the neighborhood of the throat. Luckily, we can dispense
with the twist immediately. Indeed, the twist equation \cite{Wald}
is a simple, first-order linear differential equation:
\begin{equation}
\frac{d \omega_{ba}}{d u} = -\theta \omega_{ba}
-2\sigma^c_{[a}\omega_{b]c},
\end{equation} 
whose exact solution (if somewhat formal in appearance) is
\begin{equation}
\omega_{ab}(u) = \exp\left(-\int_0^u \theta(s) ds \right) \;
{\cal U}_a{}^c(u) \; {\cal U}_b{}^d(u)\, \omega_{cd}(0),
\end{equation}
where the quantity ${\cal U}(u)$ denotes the path-ordered exponential
\begin{equation}
{\cal U}_a{}^c(u) = {\cal P}\exp\left(-\int_0^u \sigma \; ds\right){}_a{}^c.
\end{equation}
So, an initially hypersurface orthogonal congruence remains twist-free
everywhere, both on and off the throat:  $\omega_{ba}(0) = 0
\Rightarrow \omega_{ba}(u) = 0$.  Then the equation
$
\frac{d\theta}{d u} + \frac{1}{2}\theta^2 +
\sigma^{ab}\sigma_{ab} = -8\pi T_{ab}\; l^a l^b,
$
is seen to be valid for all $u$. Coming back to simply-flared
throats, we have two pieces of information regarding the expansion:
namely that $\theta(0) = 0$ and $(d \theta(u)/d u)_{u=0} \geq 0$,
so that if we expand $\theta$ in a neighborhood of the throat 
then we have that
$
\frac{d \theta(u)}{d u} = 
\left. \frac{d \theta(u)}{d u}\right|_{u=0} + O(u),
$
so over each point $x$ on the throat, there exists a finite range
in affine parameter $u \in (0,u^*_x)$ for which $\frac{d \theta(u)}{d
u} \geq 0$. Since both $\theta^2$ and $\sigma^{ab}\sigma_{ab}$ are
positive semi-definite, we conclude that the stress-energy is either
violating, or on the verge of violating, the NEC along the partial
null curve $\{x\}\times(0,u^*_x)$ based at $x$.  If the throat is
of the strongly-flared variety, then we see that the NEC is definitely
violated at least over some finite regions surrounding the throat:
$\bigcup_x \{x\}\times(0,u^*_x)$, and including the base points
$x$.  For strongly anti-trapped surfaces, the NEC is violated everywhere
in a finite region surrounding the entire throat, and including
the throat itself.

\section{Discussion}

The familiar flare-out property characterizing wormholes is manifested in 
the properties of
light rays (null geodesics) that traverse a wormhole: bundles of
light rays that enter the wormhole at one mouth and exit from the
other must have cross-sectional area that first decreases, reaching
a true minimum at the throat, and then increases. These properties
can be quantified precisely in terms of the expansion $\theta_\pm$
of the (future-directed) null geodesics together with its derivative
$d \theta_{\pm}/d u_{\pm}$, where all quantities are evaluated at
the two-dimensional spatial hypersurface comprising the throat.
Strictly speaking, this flaring-out behavior of the outgoing null
geodesics ($l_{+}$) defines one throat: the ``outgoing'' throat.
But one can also ask for the flaring-out property to be manifested
in the propagation of the set of ingoing null geodesics ($l_{-}$)
as they traverse the wormhole, and this leads one to define a
second, or ``ingoing'' throat. In general, these two throats need
not be identical 
(which can give rise to interesting causal properties \cite{Hochberg-Visser2}), 
but for the static limit they do coalesce and
are indistinguishable.

The flaring-out property implies that all wormhole throats are in
fact {\em anti-trapped} surfaces, an identification that was
anticipated some time ago by Page~\cite{Page}. With this definition
and using the Raychaudhuri equation, we are able to place rigorous
constraints on the Ricci tensor and the stress-energy tensor at
the throat(s) of the wormhole as well as in the regions near the
throat(s). We find, as expected, that wormhole throats generically
violate the null energy condition and we have provided rigorous
results regarding this matter. This should now settle the issue
of energy condition violations for wormholes.

Until recently, the nature of the energy-condition violations associated with
wormhole throats has led numerous authors to try to find ways of
evading or minimizing the violations.  Most attempts to do so focus
on alternative gravity theories in which one may be able to force
the extra degrees of freedom to absorb the energy-condition violations
(some of these scenarios are discussed in~\cite{Visser-Hochberg},
see also \cite{Brans-Dicke-Papers,Kar3}). But the energy condition
violations are still always present, for sweeping the energy condition
violations into a particular sector does not 
make the ``problem''
go away.  
More recently it has been realized that time-dependence lets one
move the energy condition violating regions around in {\em
time}~\cite{Kar1,Kar2,Wang-Letelier,Kim,Time-dependent}.  However, temporary
suspension of the violation of the NEC at a time-dependent throat
also leads to a simultaneous obliteration of the flare-out property
of the throat itself \cite{Hochberg-Visser3}, so this 
strategy ends up destroying the throat
and nothing is to be gained. (See also~\cite{Visser-Hochberg}.) 
It is crucial to note that we have
defined flare-out in terms of the expansion properties of light
rays at the throat and {\em not} in terms of so-called ``shape'' functions
or embedding diagrams.  While the latter can certainly be used
without risk\footnote{But even the static case requires that due care be exercised.
By the Whitney embedding theorem \cite{Guilleman}, a subset
$X \subset {\bf R}^n$ embeds in an ${\bf R}^{2n}$. So we should expect a static
throat, which is
two dimensional, to embed in an ${\bf R}^4$. The fact that embeddings of most of
the static
wormholes studied so far can be carried out in an ${\bf R}^3$ is due to the
highly symmetric nature of the wormholes chosen for study. A counterexample
is provided by the Klein bottle, which can be {\em visualized} but 
not embedded in ${\bf R}^3$, where
it self-intersects.} 
for detecting flare-out in static wormholes, they are
at best misleading if applied to dynamic wormholes. This is simply
because the embedding of a wormhole spacetime requires selecting
and lifting out a particular time-slice and embedding this
instantaneous spatial three-geometry in a flat Euclidean ${\bf
R}^n$. For a static wormhole, any constant time-slice will suffice,
and if the embedded surface is flared-out in the spatial direction
orthogonal to the throat, then it is flared-out in spacetime as
well. But if the wormhole is dynamic, flare-out in the spatial
direction does not imply flare-out in the {\em null} directions
orthogonal to the throat.

\clearpage
\addcontentsline{toc}{section}{Index}
\flushbottom
\printindex

\end{document}